# Local lattice distortions vs. structural phase transition in NdFeAsO$_{1-x}$F$_x$

M. Calamiotou[a,*], D. Lampakis[b], N. D. Zhigadlo[c], S. Katrych[c,+], J. Karpinski[c,+], A. Fitch[d], P. Tsiaklagkanos[e], and E. Liarokapis[e]

[a] *Solid State Physics Department, Faculty of Physics, University of Athens, GR-15784 Athens, Greece*

[b]*TEI Larissa, GR 41334 Larissa, Greece*
[c]*Laboratory for Solid State Physics, ETH Zurich, 8093 Zurich, Switzerland*
+ *Present address: Institute of Condensed Matter Physics, EPFL, 1015 Lausanne, Switzerland*
[d] *ESRF, The European Synchrotron,71 Avenue des Martyrs 38000, Grenoble, France*
[e]*Department of Physics, National Technical University of Athens, GR15780, Athens, Greece*

## ABSTRACT

The lattice properties at low temperatures of two samples of NdFeAsO$_{1-x}$F$_x$ (x=0.05 and 0.25) have been examined in order to investigate possible structural phase transition that may occur in the optimally doped superconducting sample with respect to the non-superconducting low-F concentration compound. In order to detect small modifications in the ion displacements with temperature micro-Raman and high resolution synchrotron powder diffraction measurements were carried out. No increase of the width of the (220) or (322) tetragonal diffraction peaks and microstrains could be found in the superconducting sample from synchrotron XRD measurements. On the other hand, the atomic displacement parameters deviate from the expected behavior, in agreement with modifications in the phonon width, as obtained by Raman scattering. These deviations occur around 150 K for both F dopings, with distinct differences among the two compounds, i.e., they decrease at low doping and increase for the superconducting sample. The data do not support a hidden phase transition to an orthorhombic phase in the superconducting compound, but point to an isostructural lattice deformation. Based on the absence of magnetic effects in this temperature range for the superconducting sample, we attribute the observed lattice anomalies to the formation of local lattice distortions that, being screened by the carriers, can only acquire long-range coherence by means of a structural phase transition at low doping levels.

[*]Corresponding author: mcalam@phys.uoa.gr

## 1. Introduction

Among the iron pnictides, the Ln1111-series (with Ln being a lanthanide) exhibits a high superconducting (sc) transition temperature and the two phase transitions (structural from tetragonal to orthorhombic at $T_s$ and antiferromagnetic at $T_N$ with a spin-density wave (SDW)) occur at distinct temperatures. As in other iron pnictides, upon doping, both transition temperatures decrease at the onset of superconductivity. Although there is a general consensus about the structural phase transitions of the undoped compounds [1] and those at low doping [2,3], it has been proposed that the orthorhombic phase appears even in the almost optimally doped Sm1111 compound [4]. That was based on a slight broadening of the tetragonal (110) diffraction peak in the synchrotron XRD patterns attributed [4] to an orthorhombic phase, stable even at optimal doping with the highest transition temperature ($T_c$), as in the hole-doped compounds $Nd_{1-x}Sr_xFeAsO$ [5]. Before this finding, there were indications about anomalous behavior at temperatures above 150 K from certain measurements as e.g., phonon and lattice modifications (in oxygen deficient Nd1111 compound [6,7]), resistivity anomaly in the parent 1111 compounds [8], NMR studies (in $SmFeAsO_{1-x}F_x$) [9], and the photoexcited carriers of superconducting $SmFe_{0.93}Co_{0.07}AsO$ single crystal that revealed an electronic nematicity around 170 K [10]. All these data show that even in the optimally doped 1111 compounds the electron-lattice system is modified at relatively high temperatures.

Based on our previous findings of lattice anomalies at 180 K in the $NdFeAsO_{0.85}$ (Nd085) compound [7] and the assumption of a structural phase transition for the analogous superconducting compound $SmFeAsO_{1-x}F_x$ [4], we carried out detailed low temperature high resolution synchrotron diffraction measurements on two Nd1111 compounds, one with low F doping and non-sc and another with close to optimal doping showing the highest $T_c$. The idea was: (1) to check whether the data for the high F doped superconducting sample can be compared with those of $NdFeAsO_{0.85}$ ($T_c$ = 53.5 K) in order to study the effect of the different dopants with the same $T_c$ on the observed structural modifications [7]; (2) To compare the results from the high doping compound with those of $SmFeAsO_{1-x}F_x$ [4] and determine whether the effect of a structural phase transition is peculiar to Sm1111 system and it does not appear in

$NdFeAsO_{1-x}F_x$. In addition, we wanted to trace the structural anomaly at low doping level, where the compound is non-sc and shows a clear structural phase transition.

In the following we present experimental data showing that lattice anomalies are present at high temperature in both concentrations, but evolve upon cooling differently in the two doping levels. These anomalies create a long range structural phase transition for low F doping, while at high doping level they seem to remain uncorrelated, apparently due to the long-range screening of the extra carriers [11].

**2. Experimental**

Two polycrystalline samples of $NdFeAsO_{1-x}F_x$ with nominal F concentrations x = 0.05 and 0.25 were prepared as described in Ref.[12]. The sample with x = 0.05 exhibits no superconducting transition but a magnetic ordering at $T_N$ = 40 K [12] while the x = 0.25 sample has a $T_c$ of 51 K with no relevant traces of any magnetic order [12]. Therefore, these two samples can be considered as representatives of the non-superconducting and the superconducting regions of the phase diagram of $NdFeAsO_{1-x}F_x$ system, respectively.

For the low temperature micro-Raman measurements we used a Jobin-Yvon triple spectrometer, equipped with a microscope (×100 magnification) and a liquid $N_2$ cooled charge coupled device (CCD) detector. For the excitation the 514.5 nm line from an $Ar^+$ laser was employed at low power (<0.1mW outside the cryostat). Micro crystallites were examined with typical accumulation times ~1-2 hours depending on the temperature, which was set in the range 77-300 K using the Linkam THMS600 cryostat. Due to the low laser power used, the local heating from the beam is expected to be small and the estimated temperature increase above nominal is less than 10 K at the laser spot.

High-resolution synchrotron X-ray powder diffraction (SXRPD) data have been collected on the beamline ID31 at the European Synchrotron Radiation Facility (ESRF), Grenoble, France using the experimental set-up described in Ref. [13]. In order to reduce absorption, a short wavelength (λ = 0.39996 Å and λ = 0.399787 Å for data collection for the x = 0.25 and x = 0.05 sample respectively) was selected with a double-crystal Si (111) monochromator and a Si NIST standard for calibration. The samples have been placed in a 0.6 mm diameter borosilicate glass capillary and cooled down to 10 K with an exchange gas continuous liquid-helium flow cryostat as described in Ref. [7]. A cryostream cooler has been used for data collection between 90-285 K from the x = 0.05 sample. High statistics high resolution diffraction patterns ($2\theta = 1\text{-}55^0$, d spacing 22.9Å-0.43 Å, step $0.002^0$ for the x = 0.05 and $0.004^0$ for the x = 0.25 sample respectively) have been collected at each temperature with the variable counting time (VCT) procedure to increase the statistics at high q values for the refinement of atomic displacement parameters (ADP). All data have been normalized for incident beam decay by a monitor viewing a scatter foil.

## 3. Results

Typical Raman spectra for both concentrations and selected temperatures are shown in Figure 1. The increase of the phonon width in the x = 0.05 sample (Fig. 1, left) for the T = 145 K and higher temperature spectra is obvious. On the other hand, in the other concentration the phonon width does not vary abruptly with temperature (Fig. 1, right). The dependence of the phonon characteristics (frequency and width) on temperature is presented in Figures 2-3, where data from the $NdFeAsO_{0.85}$ oxygen deficient compound (with $T_c = 53.5$ K) have been included for comparison. The frequency of the modes slightly increases with decreasing temperature as expected and appears to remain roughly constant below some temperature around 150 K. The width of all modes for the superconducting sample (x = 0.25) increases slightly with lowering the temperature, contrary to the expected behavior. The width of the Nd mode with the better statistics follows the same trend with the results of $NdFeAsO_{0.85}$ compound (Fig. 3). For the x = 0.05 compound the width is modified substantially and rather suddenly below ~150K (Fig.3). This is the temperature range where the parent compound undergoes a structural phase transition [1,3] and where resistivity has shown an abnormal modification with temperature [8]. Our XRD structural measurements have also indicated lattice modifications close to this temperature.

Data analysis of the XRD diffractograms on the two $NdFeAsO_{1-x}F_x$ compounds has been performed with the GSAS+EXPGUI suite of Rietveld analysis programs [14]. The peak profile was modeled by a pseudo-Voigt function corrected for asymmetry owing to axial divergence. The peak broadening has been modeled in the refinement with the anisotropic microstrain description of Stephens [15] (profile # 4 in GSAS). The diffraction patterns of the x = 0.05 sample have been fitted using the tetragonal

(*P4/nmm*) structure in the temperature region 290 K - 70 K and the orthorhombic (*Cmma*) structure in the temperature region 60 K - 20 K. For this sample, a clear splitting of the $(220)_T$ tetragonal reflection into the $(400)_O$ and $(040)_O$ orthorhombic reflections is observed at 60 K, unambiguously indicating an orthorhombic structure. At 70 K only a peak broadening can be detected and the fit using a tetragonal structure model incorporating an anisotropic microstrains description rather than using an orthorhombic one results in slightly better R values. The diffraction patterns of the x = 0.25 sample at all temperatures down to 10K reveal *no splitting or any sign of broadening of the corresponding reflections*. Our high statistics data allow to probe this with the $(322)_T$ tetragonal reflection at a high q value ($2\theta \approx 21.7^0$), which splits to the $(512)_O$ and $(152)_O$ in the presence of an orthorhombic distortion of the average cell (shown in inset of Fig. 4 for RT and 10 K). Therefore, for the x = 0.25 sample the patterns have been fitted with the tetragonal (*P4/nmm*) structure in the whole temperature region down to 10 K and the possibility of a T-O phase transition in this compound will be examined in what follows from the evolution of anisotropic microstrains incorporated in the profile refinement.

Figure 4 shows the experimental diffraction pattern of the x = 0.25 (at 10 K) sample, together with the results of the Rietveld refinement. The Nd and As coordinates, the isotropic atomic displacements parameters (ADP), $U_{iso}$, of all atoms and the corresponding Stephens profile parameters [14,15] have been sequentially refined. Oxygen and fluorine occupancies have been fixed to the nominal values justified by the agreement of the amount of relative volume reduction due to fluorine replacement to that reported for the $NdFeAsO_{1-x}F_x$ system. [3] In the refinements small amounts of impurity phases have been included (2.8%wt NdOF in x = 0.05 and 7.94(2)%wt NdOF, 5.63(2)%wt NdAs, 6.80(4)%wt FeAs in the x = 0.25 sample).

The temperature dependence of the lattice constants is shown Fig. 5 for both compounds. For the x = 0.05 sample, the $a_O/\sqrt{2}$ and $b_O/\sqrt{2}$ values are plotted in the O-structure range. The relative volume contraction due to the smaller fluorine atoms replacing oxygen in the x = 0.25 sample at room temperature amounts 0.96% in agreement to that obtained from Ref. [3]. For the x = 0.05 sample a change of slope in the temperature dependence of the c-axis is observed between 70 K and 60 K. This feature persists when we refined the 70 K data with an orthorhombic cell. A similar effect (different slope for c below 140 K), as well as an anomaly in the course of the atomic displacement parameter of Fe atom has been observed for single crystal GdFeAsO well above the structural transition ($T_s$ = 111 K) [16]. A change of slope in the temperature dependence of the *c*-axis through the structural transition has been also reported for the LaFeAsO compound, suggesting that the structural distortion in LaFeAsO occurs over a wide temperature range, but includes a "sharp" anomaly as well [17]. These lattice anomalies observed in a temperature region well above the transition temperature appears to be a common characteristic of the parent Ln1111 compounds. On the contrary, the dependence of c-axis on temperature is smooth for the x = 0.25 sample to the lowest temperature studied (10K), except of a small change in slope close to $T_c$. The a-axis of the x = 0.25 compound shows some small anomalies that begin around 160 K and end at $T_c$ (Fig. 5a). Similar but more pronounced anomalies have been observed in the NdFeAsO$_{0.85}$ compound [7].

The most important part of the structure of the Ln1111 compounds is the superconducting Fe-As layer. The contraction of the Fe-As layer thickness (normalized to the high temperature value) along the *c*-axis (see Fig. 4a in Ref.[18]) upon cooling is shown in Figure 6. Both compounds exhibit the same relative contraction down to 150 K. While the superconducting compound exhibits a smooth contraction down to 10 K, a change of slope at T ~ 150 K and a discontinuity at T = 70 K is observed for the x = 0.05 composition. These results of the average structure do not support the hypothesis of a long-range T-O transition in the superconducting sample. It is worth to notice that all the above modifications persist when we fit the data with the orthorhombic structure in the temperature region 70 K - 150 K. Similar effects previously reported for the LaFeAsO compound have been correlated with the onset of the magnetic order [19]. Since in the NdFeAsO$_{0.95}$F$_{0.05}$ sample $T_N$ (40 K) is well below the observed discontinuity, we can assume that the observed relaxation of lattice distortions in the non-superconducting sample is mostly related with the T-O transition. Furthermore, no sign of any magnetic order has been detected on the same compounds at high temperatures [12], excluding the possibility of any direct involvement of the spin ordering in these lattice effects.

## 4. Discussion

X-ray diffraction data collected at a synchrotron source can provide also microstructural information on the local scale from the broadening of diffraction peaks. Moreover, random local dynamic or static displacements from the ideal atom positions manifest in the refined atomic displacement parameters (ADP´s)-which can be refined with our high statistics data. Microstrain-type broadening of the diffraction lines in the Stephens model that we have used in the refinements is considered as a manifestation of the distribution of the lattice parameters. Although the latter can originate from different effects such as, substitutional inhomogeneity or crystal defects, local orthorhombic distortions, if present in the long range tetragonal lattice, are expected to increase the microstrains along the [110] direction with decreasing temperature and to relax in the orthorhombic structure. Figure 7 shows for both compounds the evolution of microstrains along the [110] and [001] directions calculated from the refined Stephens coefficients with temperature. Since we have investigated the evolution of the microstrains with temperature rather than their absolute values and taking into account the very small line width of a standard sample measured at ID31, the instrumental effect can be assumed to be negligibly small [20]. Size type broadening has been also found to be negligibly small. The microstrains along the [*hkl*] direction have been calculated, as described in Ref.[20], taking into account that in the tetragonal structure the diffraction width is parameterized with four refinable coefficients $S_{400}$, $S_{004}$, $S_{220}$, $S_{022}$ [14,15].

For the x = 0.05 sample upon cooling microstrains along the [110] direction (Fig. 7) start to increase at the temperature range 150-180 K, reaching a maximum increase of ~20% at 70 K, and relaxing (below 70 K) to lower values in the orthorhombic structure. On the contrary, the microstrains along the [001] direction remain almost constant from room temperature to ~150 K to slightly increase below this temperature. This suggests that there are some lattice distortions within the *ab*-plane related with the orthorhombic phase transition. Although a clear splitting of the characteristic reflections could be observed only at 60 K for x = 0.05, one could assume that the phase transition starts already around 150 K as indicated by the microstrains along the [110] direction (Fig. 7).

Concerning the microstrains of the x = 0.25 sample, in the [110] direction there is no variation contrary to the case of the x = 0.05 compound. On the other hand, in the superconducting sample an increase of microstrains is observed upon cooling along the [001] direction reaching a maximum at ~60 K with a small but noticeable relaxation below $T_c$ (Fig. 7). Our results suggest an increase of local lattice disorder only along the *c*-axis in the superconducting compound, in full agreement with the absence of any broadening of the line widths discussed before. One may also notice that microstrains are in general higher and anisotropic for the x = 0.25 sample, which could arise from the increased amount of F replacement for O. The different behavior of microstrains in the two compounds does not support a structural phase transition in the sc compound, although some local lattice distortions could happen on a nanoscale.

Our results for the NdFeAsO$_{1-x}$F$_x$ system are in contrasts with the SmFeAsO$_{1-x}$F$_x$ system where the width of the (110)$_T$ line of a x = 0.2 superconducting sample has been reported to increase by ~12% in the temperature region 100-210 K that was attributed to a T-O phase transition [4]. A structural phase transition in the SmFeAsO$_{1-x}$F$_x$ system has been also recently anticipated from $^{19}$FNMR studies [9]. Since NMR is a local probe, the modifications observed around 163 K in the relaxation times point to a local lattice distortion and it is unclear whether this acquires also a long range. Pair distribution function (PDF) studies on two Sm1111 compounds indicate also that the local structure differs significantly from the average structure determined by Bragg diffraction [21]. Finally, the study of the photoexcited carriers in a superconducting SmFe$_{0.93}$Co$_{0.07}$AsO single crystal revealed a 2-fold symmetry breaking in the tetragonal phase related to electronic nematicity around 170 K [10]. All these results point to some lattice distortions that appear in superconducting compounds at much higher temperature than $T_c$. However, the question arises whether *these distortions acquire a long range order as a structural phase transition or they remain isolated in uncorrelated domains*.

Additional information for the local lattice disorder can be obtained from the evolution with temperature of the refined isotropic atomic displacement parameters (ADPs), U$_{iso}$, of the different atoms. ADPs represent thermal atomic motion and possible static displacive disorder [22]. Assuming the simple Debye model to describe the temperature-dependant part arising from atomic vibrations, the evolution of isotropic ADPs with temperature can be expressed as $8\pi^2 U_{iso} = \frac{6h^2}{mk_B \Theta_D}\left[\frac{\varphi(x)}{x} + \frac{1}{4}\right] +$ $d^2$ with $x = \frac{\Theta_D}{T}$ and $(x) = \frac{1}{x}\int_0^x \frac{\xi}{e^\xi - 1}d\xi$, where $\Theta_D$ is the Debye temperature, *m, k$_B$, h* have their usual meanings and d$^2$ is a term introduced to account for the non-thermal disorder in the lattice [22]. The temperature dependence of the refined $U_{iso}$ values for Nd, Fe and As atoms in both compounds is shown in Figures 8 (a-c). Solid and dashed lines represent simulated typical Debye-type temperature dependence for $U_{iso}$. We have used in the calculations Debye temperatures $\Theta_D$ =355(10) K, similar to that reported for the homologous PrFeAsO compound [23], and additional displacive disorder terms $d^2$. For the As and Fe atoms small terms have been added, i.e., $d^2$ = 0.001 Å$^2$ (0.00125 Å$^2$) for the As in the x = 0.25 (x = 0.05) compound and 0.0008 Å$^2$ for the Fe atoms in the x-0.05 compound. For the Nd atoms a $d^2$ = 0.0023 Å$^2$ and $d^2$ = 0.0043 Å$^2$ term must be added for the x = 0.05 and x = 0.25 compound respectively showing that Nd atoms exhibit larger disorder effects in the fluorine rich sample.

The high temperature (T>150 K) $U_{iso}$ values of the Fe and As atoms are comparable in the two compounds following for both the anticipated Debye-type temperature decrease upon cooling. From ~ 150 K down to 10 K on the other hand, the $U_{iso}$ values of the x = 0.25 compound increase indicating that an additional *structural disorder effect* sets in (Fig. 8b, c). Something similar is observed in the $U_{iso}$ of Nd, but in that case the initial high temperature values at ambient temperature differ substantially in the two compounds, most probably due to the larger distortions induced in the Nd-O layer by the substitution of oxygen by fluorine (Fig. 8a). In order to rule out that these large $U_{iso}$ values in the sc compound originate from *apparent* displacements [22] due to the use of the tetragonal symmetry instead of an orthorhombic one in the region T≤150K, we have repeated the refinements using the Cmma symmetry (open stars in Fig. 8a-c). The evolution of $U_{iso}$ values with temperature in this case shows the same characteristics proving that the increased random local atomic displacements in the T≤150K region are an intrinsic characteristic of the sc phase *and it is not related with a hidden structural phase transition to an orthorhombic phase at high temperatures.*

For the x = 0.05 non-sc compound a change of slope at $T_{onset}$ ~ 150 K is also evident in the temperature dependence of the $U_{iso}$ values of the Fe and more pronounced of the As and Nd atoms. The temperature dependence of random local lattice distortions, expressed as ADP of the Nd atom shows the same trend with the width of the Raman active modes (Fig. 3). The same is true for the other two phonons (As and Fe), although the weakness and the proximity of these two phonon modes did not allow more accurate determination of their widths. Similar temperature dependence of the Nd mode has been found for the oxygen deficient sc compound (Fig. 3 and Ref. [24]), as well as for other F concentrations (Fig. 9). In all cases, the width remains roughly constant with a small tendency to increase below some temperature (as in the $U_{iso}$ of the same atom, Fig. 8a).

Anomalies in the Raman spectra have been also observed in the 122 compounds ($Ba_{1-x}K_xFe_2As_2$ and $Sr_{1-x}K_xFe_2As_2$), where the width changes at the structural phase transition or the SDW [25]. XRD and neutron scattering measurements on $Ba_{1-x}K_xFe_2As_2$ have shown a lattice softening at higher temperatures than the structural phase transition, which has been attributed to a weak orthorhombic distortion at the microscopic scale [26], in agreement with the modifications in the phonon characteristics [25]. It appears that in all these systems changes in the phonon characteristics unmask lattice modifications that occur at high temperatures.

The dependence of both, atomic displacement parameters and Raman mode widths, on temperature suggest that a precursor local lattice instability occur in the $NdFeAsO_{1-x}F_x$ system around 150 K for both F dopings, with distinct different temperature evolution for T<150 K among the two compounds. Instead of a pronounced increase of the $U_{iso}$ values of the sc compound, a decrease is observed for the low doping sample. This indicates that the tendency to disorder follows different paths in the two compounds, apparently from the additional doping.

A change of slope in the atomic displacement parameters may indicate the onset of a structural phase transition, which cannot be clearly resolved by a corresponding change of the refined average structure [22]. For the x=0.05 compound the observed change of slope at $T_{onset}$ ~ 150K agrees with the onset of the T-O structural phase transition at the same temperature as indicated by the results of the microstrain analysis (Fig. 7). Contrary, in the x=0.25 sc compound, our microstrain analysis does not support the hypothesis of a similar long-range T-O structural phase transition in the hole temperature region studied. Both, the increased $U_{iso}$ values (even in the analysis of an apparent orthorhombic symmetry) and the increase of the width of all Raman modes for the sc sample (x = 0.25) with lowering the temperature, indicate that in this case an isostructural phase transition to a disordered phase takes place in the presence of excess carriers with the local lattice displacements not acquiring long range coherence. Finally, one could remark that the same lattice effect is observed on the Fe-As layer and the Nd-O(F) layer indicating that increased local lattice disorder is evident in different sites of the unit cell.

The results of the PDF analysis [21] or the photoexcited carriers [10] and the comparison of the two compounds points to a scenario of local lattice instability (a local atomic displacement mechanism), which sets-in for the Nd1111 system at high temperature. In the low doping level this lattice instability leads to a structural long range T-O phase transition while in the high doping level it cannot acquire long range coherence leading only to an isostructural transition to a sc phase with non correlated local atomic displacements. To this context, the coupling of lattice with carriers may play a crucial role in the long range coherence mechanism of local atomic displacements.

Considering the origin of effective displacement of atoms from the main positions, besides the atomic vibrations, that could result in changes of the $U_{iso}$ values at $T_{onset} \sim$ 150 K, density and displacement modulations or short-and long-range displacive correlations could induce similar modifications. The former could be associated with spin or carrier modulations (SDW or CDW) while the latter could be in accordance with the stripe scenario or local orbital ordering mechanism. The sc sample, studied by muon-spin spectroscopy [12], does not show any sign of short or long-range magnetically ordered phase, while the SDW long range magnetic order transition temperature of the non-sc x = 0.05 sample is 40K. Therefore the coupling of spin density fluctuations with the lattice seems to be unlikely the origin of the lattice instability that sets in at 150K in both compounds. A CDW mechanism seems also to be mostly unlikely since, to our knowledge, no such evidence has been reported so far.

On the other hand, orbital ordering of a Jahn-Teller type distortion driven by the Coulomb repulsion has been proposed as the driving mechanism for the T-O structural phase transition (SPT) in the low doping level, while extra electrons or holes tend to diminish the Jahn-Teller effect preventing the SPT in the sc compounds [11]. Following a symmetry-mode and spontaneous strain analysis of the Ln1111 compounds, A. Martinelli proposed that a macroscopic distortion in the tetragonal lattice takes place as local microstrain fields (probably driven by the tendency of Fe orbitals to order), start to interact in a correlated or collective way [27]. Local breaking of symmetry at the Fe sites due to orbital ordering could be one of the driving mechanisms of local atomic displacements (lattice instabilities) that sets in at $T_{onset} \sim$ 150 K in both compounds (Ln = Sm or Nd), but evolves differently with temperature apparently due to the coupling of excess carriers with the lattice. But, the deviation from the simple Debye model, for both underdoped and optimally doped concentrations, seems to be smaller for Fe than for the As or Nd atoms (Fig. 8). Therefore, orbital ordering in the way assumed in Ref. [27] does not comply with our data. The possibility the observed effect to be due to the disorder introduced by the F doping can be excluded, since in that case the $U_{iso}$ values had no reason to substantially increase at low temperatures.

Concerning the SmFeAsO$_{1-x}$F$_x$ system, it is unlike that the Sm substitution for Nd could alter the above mentioned mechanism. We have tried to reveal any potential lattice anomaly by Raman spectroscopy in SmFeAsO$_{1-x}$F$_x$ and the results are presented in Fig. 10 for the parent compound and the optimally doped one. To the lowest temperature studied (77 K) no anomaly has been observed in the doped sample and its temperature dependence follows closely that of NdFeAsO$_{1-x}$F$_x$ (Figs. 3 & 9). Therefore, the structural phase transition assumed by Martinelli *et al* [4] may need further verification, since our Raman data on SmFeAsO$_{1-x}$F$_x$ do not support such hypothesis.

## 5. Conclusions

We have measured two concentrations of the NdFeAsO$_{1-x}$F$_x$ compound to investigate the structural distortions up to room temperature at low and optimal doping. The results from the average structure as well the microstrains and the FWHM of the XRD lines do not reveal any phase transition in the optimally doped sc sample, contrary to the low doping one. The analysis of the atomic displacements also show that the phase transition is absent in the optimally doped compound. Raman measurements indicate a strong similarity between the temperature dependence of the phonon widths and the local lattice distortions expressed as atomic displacements parameters. All our data indicate local lattice distortions that evolve to an orthorhombic structural phase transition only at low carrier concentration. Apparently, the excess carriers in the sc concentration screen the interaction between the local deformations and the compound remains in the tetragonal phase in the optimal doping. The absence of any sign of magnetic ordering points to a different interaction mechanism of the carriers with the lattice. On the SmFeAsO$_{1-x}$F$_x$ system we have not observed any phonon anomaly in the Raman data, which could imply that in this pnictide as well the same effect of uncorrelated local lattice distortions applies.

**Acknowledgments**

We acknowledge the European Synchrotron Radiation Facility for provision of synchrotron radiation facilities, at beamline ID31 (proposal ch3478).

**Figure Captions**

**Figure 1**. Raman spectra of the NdFeAsO$_{1-x}$F$_x$ compounds x = 0.05 (left) and x = 0.25 (right) at selected temperatures.

**Figure 2.** (Color on-line) Phonon frequency as a function of temperature of the Nd, As and Fe modes for the two NdFeAsO$_{1-x}$F$_x$ compositions. Open squares are data from the NdFeAsO$_{0.85}$ oxygen deficient compound.

**Figure 3.** (Color on-line) Phonon width as a function of temperature of the Nd, As and Fe modes for the two NdFeAsO$_{1-x}$F$_x$ compounds. Open squares are data from the NdFeAsO$_{0.85}$ oxygen deficient compound. Lines are guides to the eye.

**Figure 4**. (Color on-line) Experimental (red crosses), calculated (continuous line) SXRPD patterns, and their difference (bottom line). Bars indicate Bragg peak positions for NdFeAsO$_{1-x}$F$_x$ (x=0.25). The three lower rows of bars indicate the Bragg positions of the impurity phases NdOF, NdAs and FeAs, respectively. The inset shows the profiles of the (322)$_T$ reflection at 10K and RT with peaks being shifted so as to be superimposed.

**Figure 5.** (Color on-line) Temperature dependence of (a) the *a*-, *b*-axis and (b) the *c*-axis for the NdFeAsO$_{1-x}$F$_x$ (x = 0.05, 0.25) compounds.

**Figure 6.** (Color on-line) Temperature dependence of the Fe-As layer thickness along the *c*-axis (normalized to its high temperature value) for the x = 0.05 and x = 0.25 compounds.

**Figure 7**. (Color on-line) Evolution of microstrains with temperature along different directions for both compounds.

**Figure 8**. (Color on-line) Temperature dependence of the U$_{iso}$ of Nd (a), Fe (b), and As (c) atoms for both compounds. Solid lines represent simulated typical Debye-type temperature dependence for U$_{iso}$ and open stars present refined values using an orthorhombic symmetry (see text).

**Figure 9.** (Color on-line) Nd phonon width as a function of temperature for the NdFeAsO$_{1-x}$F$_x$ compounds with different F concentrations. Solid lines are guides to the eye.

**Figure 10**. (Color on-line) Temperature dependence of the Sm and As mode phonon width for the SmFeAsO$_{1-x}$F$_x$ compounds. Solid lines are guides to the eye.

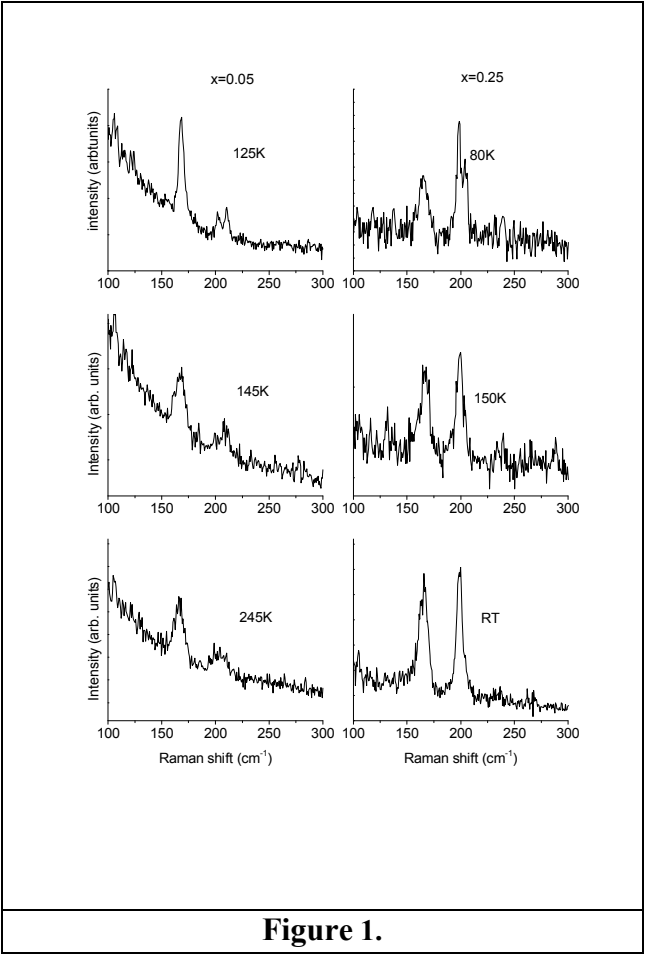

**Figure 1.**

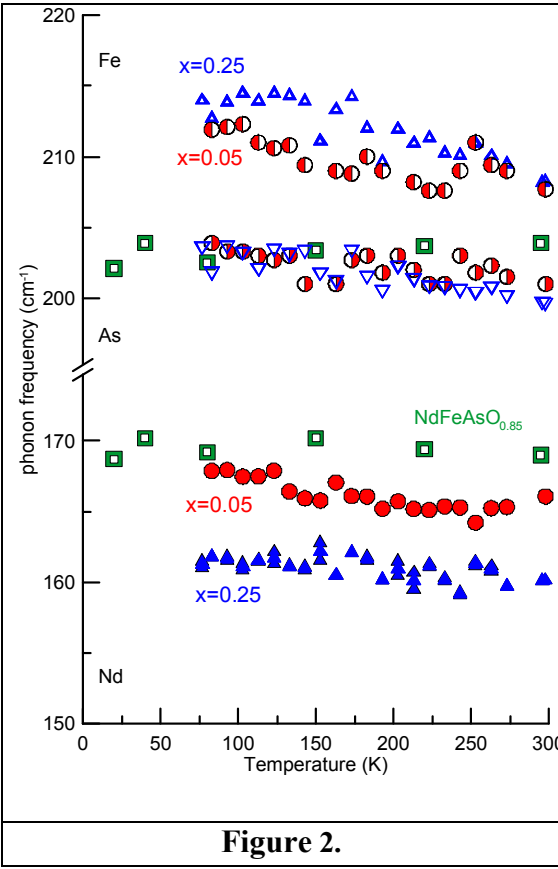

**Figure 2.**

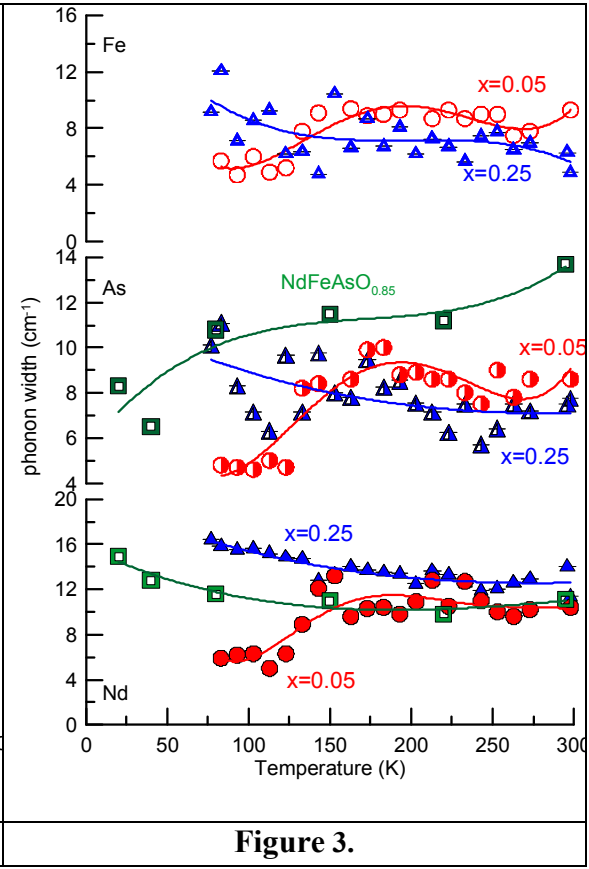

**Figure 3.**

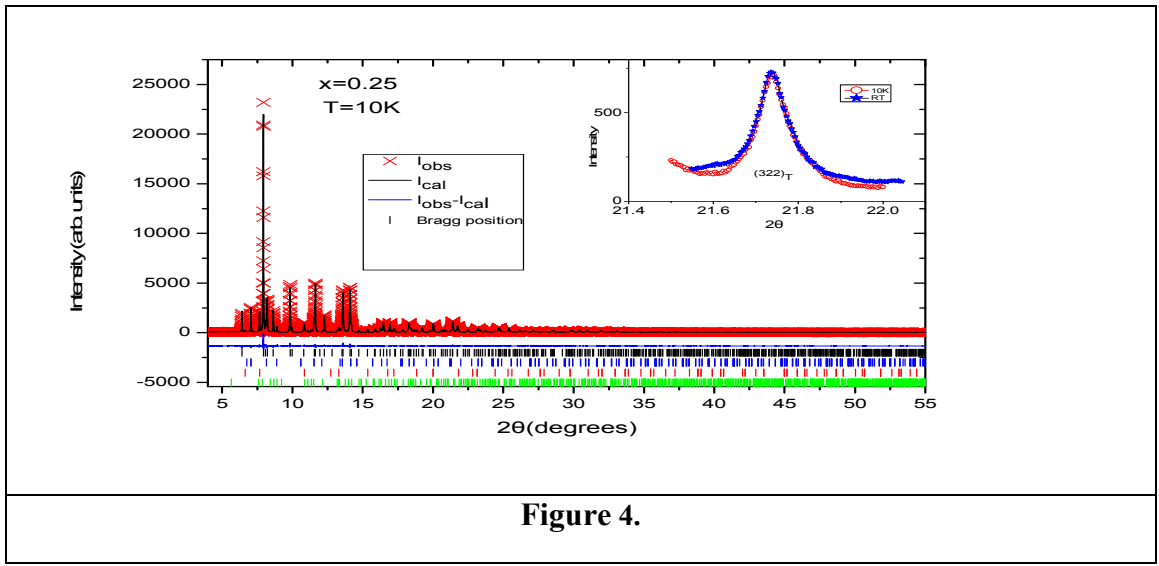

**Figure 4.**

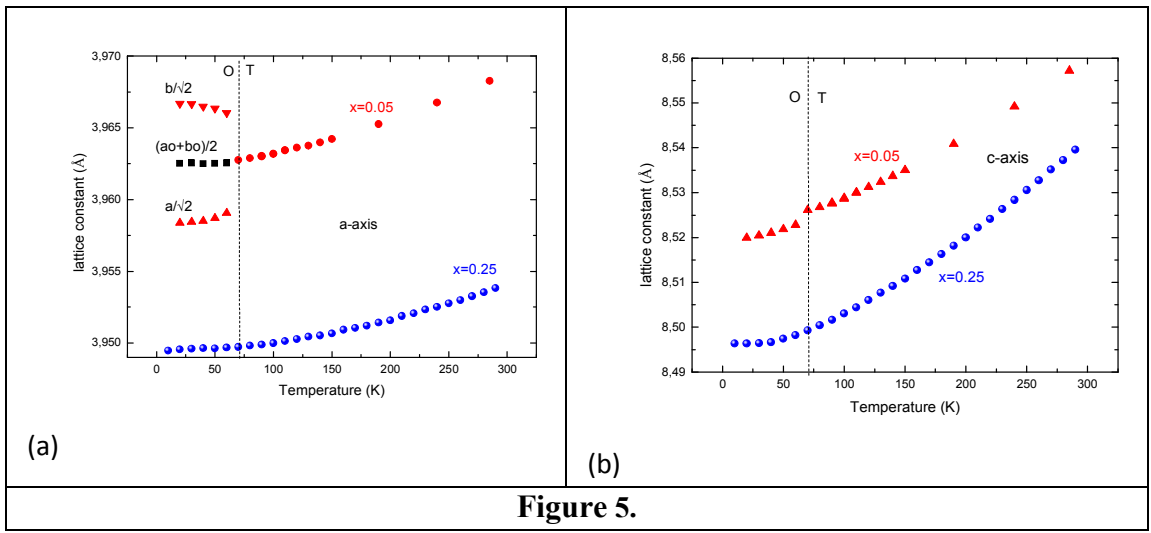

(a)   (b)

**Figure 5.**

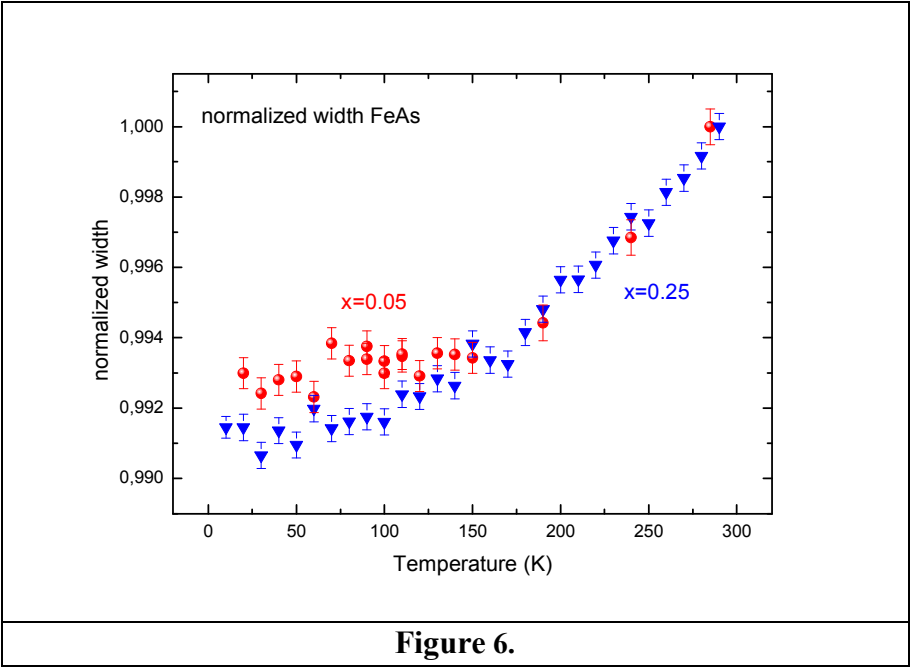

**Figure 6.**

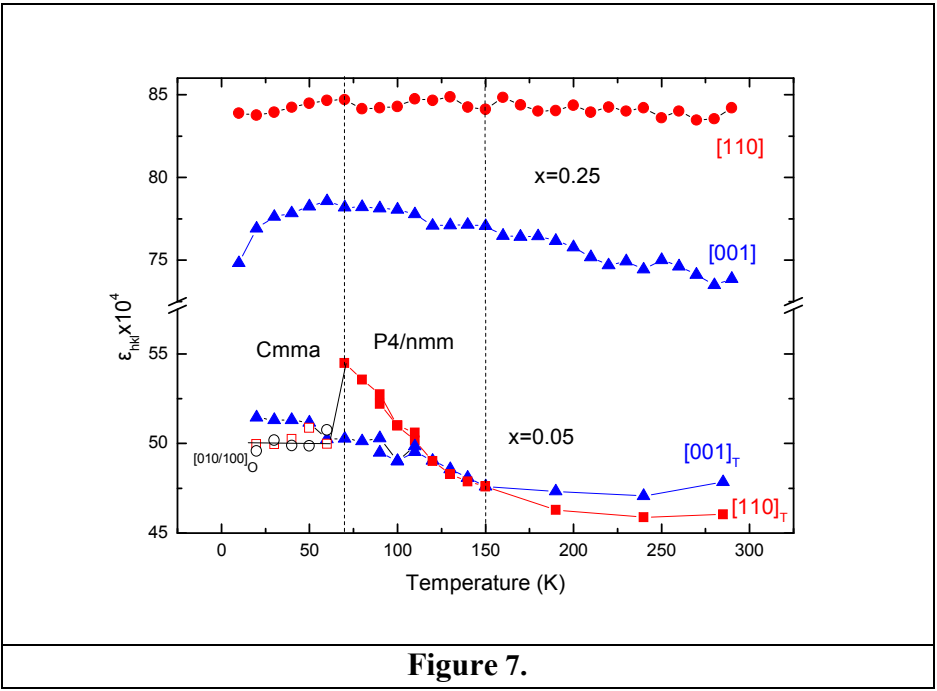

**Figure 7.**

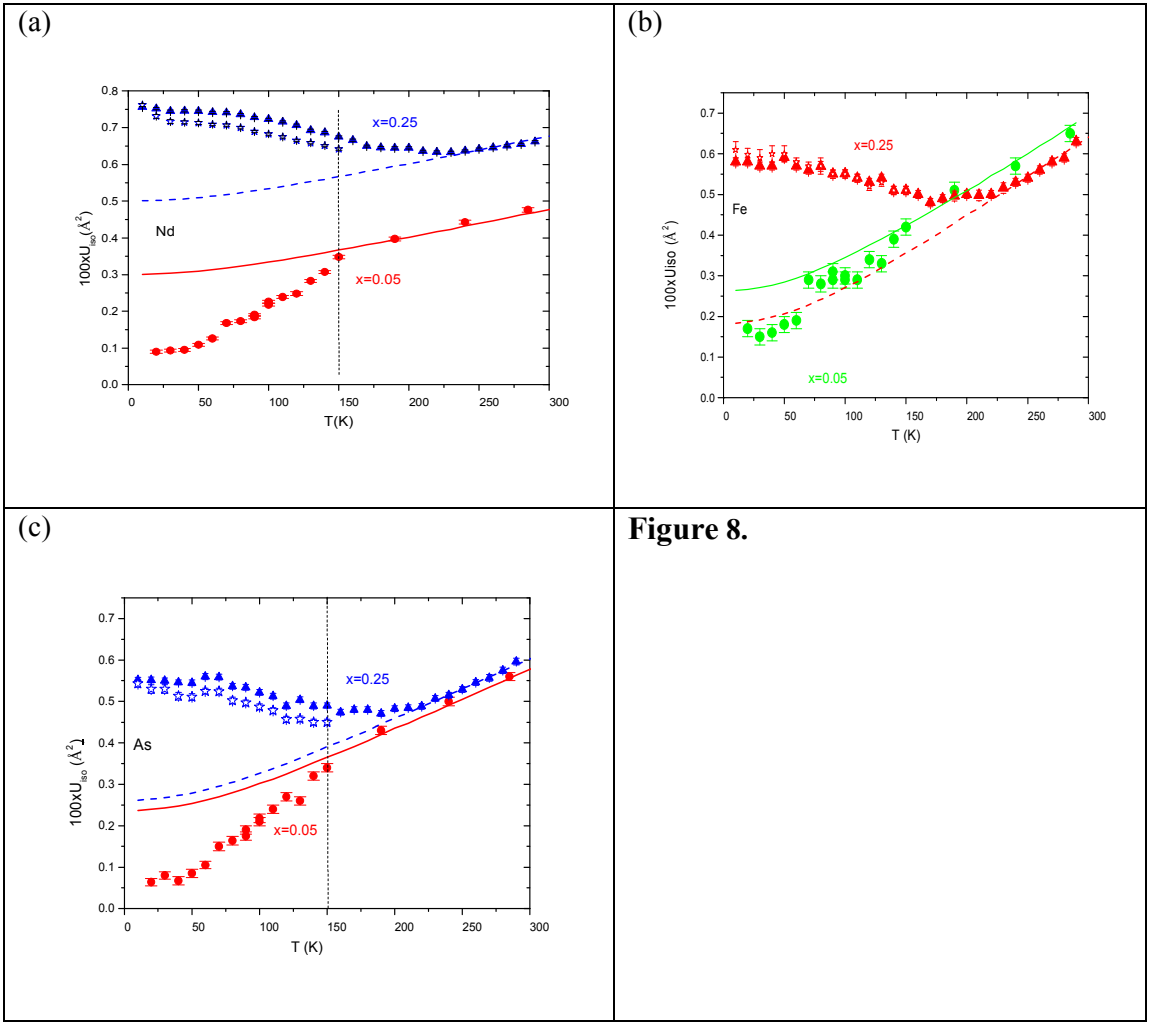

Figure 8.

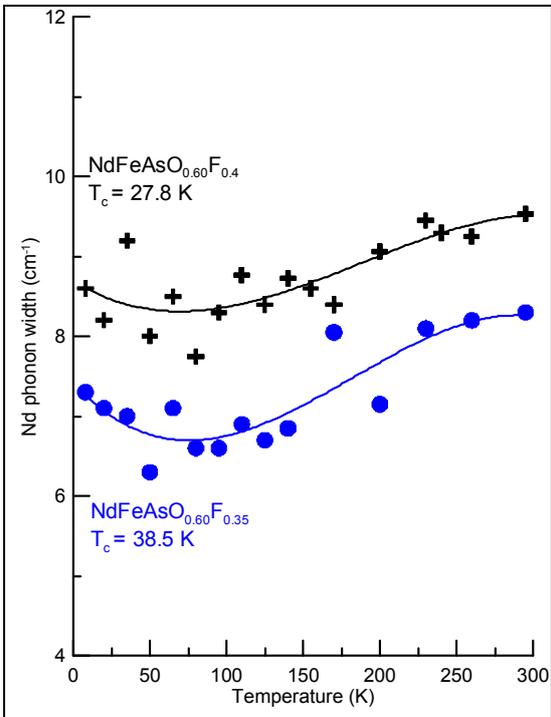

Figure 9.

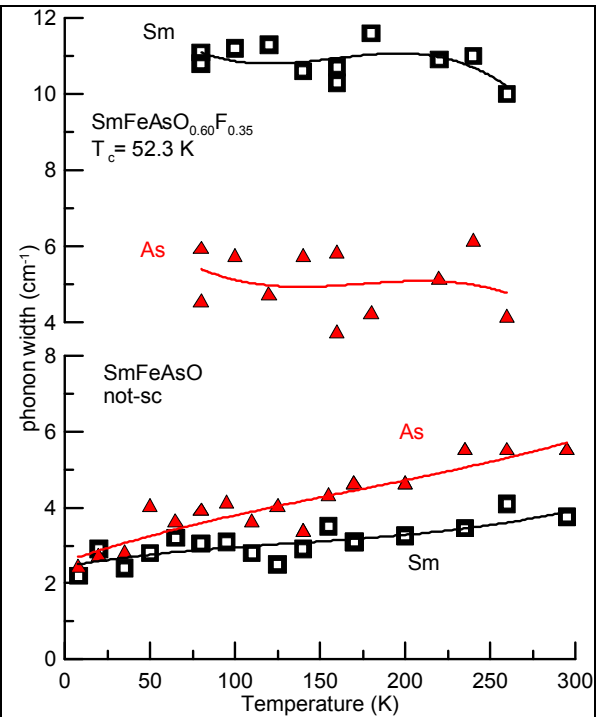

Figure 10.